\documentstyle[multicol,aps,prb,epsf,graphics,psfig]{revtex}

\begin{document}
\def\be{\begin{equation}}
\def\ee{\end{equation}}
\def\bea{\begin{eqnarray}}
\def\eea{\end{eqnarray}}

\draft
\preprint{}

\title{Coherent motion in the interaction model of cold glasses}

\author{Abdollah Langari}

\address{
Institute for Advanced Studies in Basic Sciences,
Zanjan 45195-159, Iran}
\address{
Max-Planck-Institut f\"ur Physik komplexer Systeme,
N\"othnitzer Strasse 38,
D-01187 Dresden, Germany }
\date{\today}

\maketitle

\begin{abstract}
\leftskip 2cm
\rightskip 2cm
We have studied the collective phenomena of multicomponent glasses
at ultra low temperatures [Strehlow, et. al, Ref.\onlinecite{strehlow}] by taking
into account the proper interaction between tunneling centers.
We have considered both double and triple well potentials with different
types of interactions. We show that a phase with coherent motion
appears for a range of parameters when the path of tunneling
is coursed by an interaction of the XY type, while the usual Ising like
interaction does not lead to the expected collective phenomena.
In the phase of coherent motion,
the dipole moment and the low-energy levels oscillate with a frequency
proportional to the number of tunneling centers in the system.
Simultaneous level crossing occurs between the ground and first
excited states.
The effects of long-range interactions
and also of random couplings have been also studied for a one-
and two-dimensional array of tunneling centers. We find that
long-range interactions do not affect the coherent motion, while
a wide distribution of random couplings destroys the collective effects.
\end{abstract}



\begin{multicols}{2}
\section{Introduction}

\begin{figure}
\epsfxsize=7.5cm \epsfysize=7.5cm {\epsffile{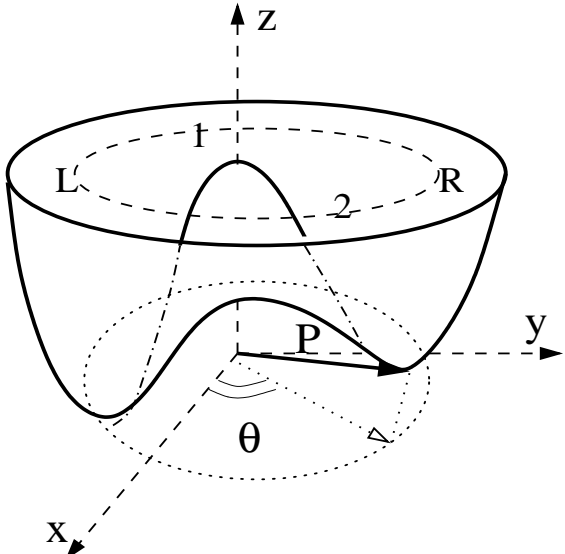}}
\caption{
A double well potential in three dimensional space. The wells are
marked by L and R which are separated by a barrier. The tunneling may
happen through this barrier via path 1 or 2 . The position of
a charged particle in the potential defines the dipole moment {\bf P}.
The projection of {\bf P} onto the x-y plane is described by
the angle $\theta$ measured
from the x-axis.
}
\label{fig1}
\end{figure}

At low temperatures glasses exhibit surprising phenomena,
which are interesting both from a theoretical as well as
from an experimental
point of view. These properties have been attributed to
the low-lying excitations which appear in almost all amorphous and disordered
solids.\cite{esquinazi} The most simple and successful model to
explain many properties of glasses is the tunneling
model (TM).\cite{phillips,anderson}
In this model the excitations are
described phenomenologically as a tunneling system.
One may consider a double-well potential to be a tunneling  system where
an entity (an atom, group of atoms, ...) tunnels
between the wells. At low temperatures only the lowest lying state
of each well is relevant, so the model is effectively described
by a two-dimensional Hilbert space where the basis kets represent the
ground states of each of the well. Using the Pauli matrices, the Hamiltonian
of an isolated two-level system (TLS) is given by
\be
H_0=\frac{1}{2}(\Delta \sigma^z + \Delta_0 \sigma^x).
\label{h0}
\ee
The eigenstates $|+\rangle$, $|-\rangle$ of $\sigma^z$ refer to the
particle being in the right (R) and left (L) well, respectively
(see Fig. (\ref{fig1})).
$\Delta$ is the asymmetry energy and $\Delta_0$ is the tunneling
matrix element. The analogy of $H_0$ with a spin-1/2 particle
in a magnetic field can be used to explain the dynamics of TLS's in glasses.
The interactions with acoustic and electric fields are usually treated as
weak perturbations which enables the TLS model to explain successfully
many of the
anomalous thermal, acoustic and dielectirc properties of glasses.

Although the isolated TLS model works very well in many cases, there
are some phenomena which can not be described only by isolated TLS's.\cite{enss}
In these cases it is expected that the interactions between the TLS's will
resolve the problem.\cite{yu,kassner}
As an example, one may consider the effects
of interactions in the two-level tunneling defect crystals, KCL:Li.\cite{wurger}
In this system the density of defect ions (Li$^+$) is a tunable parameter.
At low density, the distance of tunneling centers is fairly large, so
the interaction does not affect the behavior of the isolated TLS's.
Upon increasing the
density, dipole-dipole interactions between the TLS's become
important, and drive the system into a glassy state.\cite{kuhn}

Recently experiments on multicomponent glasses show novel phenomena at
ultralow temperatures.\cite{strehlow} By decreasing the temperature down
to 5 mK the dielectric constant $\epsilon$ responds linearly
to a small magnetic field
of the order of 10 $\mu$T.
In general, glasses are properly
assumed to be linear dielectrics, so that $\epsilon$ depends quadratically
on the electromagnetic field.\cite{reijntjes}
Although the dependence on the
magnetic field may come from nonlinear effects, the observed magnetic
field dependence of $\epsilon$ in multicomponent
glasses at very low temperatures
is completely different in nature from the magnetoeffect in nonlinear
dielectrics,\cite{strehlow} and can not be derived from thermodynamics
by assuming glasses as simple magnetizable dielectrics.

The first theoretical approach addressing this behaviour
considered the Aharanov-Bohm effect upon a charged particle in a
three dimensional double-well potential.\cite{kettemann} In this
approach, as explained briefly in the next section, the effect of
magnetic field enters in the form of a flux-dependent hopping parameter in
the standard TLS. Then the isolated TLS of Eq.(\ref{h0}) is studied
with $\Delta_0\equiv\Delta_0(\phi)$. To fit the experimental
electric permittivity data (Fig. 3 in Ref.\onlinecite{kettemann}) the charge
entity in the TLS should be Q$\approx10^5|e|$, where $|e|$ is the magnitude
of the electron charge. The large value of Q is explained
by assuming  {\it coherent motion} of the charged particles in all of the
TLS's on a mesoscopic scale. The coherent motion takes place when the
interactions between the TLS's become important at ultralow
temperatures. In this phase one may treat the motion of all of the
TLS's in terms  of a single TLS with a charge Q which
is equal to the sum of
all of the charges. Although this  model effectively describes the
magnetic field dependence of the electric permittivity, the type of
interaction which may lead to coherent motion over the mesoscopic size
of TLS remained open.

Our goal in this paper is to explain in a more quantitative way the
collective effect of TLS's in the presence of a magnetic field by
taking into account the interactions of the TLS's.
In this approach we introduce
the magnetic field in the flux dependent hopping elements,
while the interaction term traces the path of the tunneling process.
The coherent motion of the particles is simulated by probing
the relative motion of each coupled TLS.
We suppose that our model is the original interacting model
of the effective TLS containing the renormalized parameters
introduced in Ref. \onlinecite{kettemann}.
We will study the effects of long-range (l.r.) interactions  and also
of random couplings using our model in order to justify
our contention that it
represents the collective phenomenon of cold glasses properly.

The outline of this paper is as follows : In the next section we will
briefly explain the effective model of TLS's in which the magnetic field
is included as a flux-dependent tunneling matrix
element.
In Sec. III the double and triple well model are studied in the
presence of an Ising-like interaction. We introduce
the XY-like interaction
in Sec. IV. In this model, it is possible to trace the
tunneling path  in each
TLS and hence to simulate the coherent motion. We will study
the effect of long-range  interactions and random couplings on
the behavior of our model in different subsections. In these studies
we have used an exact diagonalization method to obtain the physical quantities
from our model restricted to one or two dimensional lattices.
We finally present our conclusions in Sec. V.

\section{ Effective model of TLS}
We start first by explaining the effective
model of an isolated  TLS.\cite{kettemann}
The basic idea in this approach is to consider the Aharanov-Bohm effect upon
a charged particle in a two-dimensional double well potential.
The real space representation of the potential might look similar to
a Mexician hat,
as shown in Fig. \ref{fig1}. In this figure the solid lines
defines a potential with two minima separated by a finite barrier.
The two minima are labeled  L and  R.
A particle with charge $Q$ bounded in this potential may tunnel from
one well (L) to the other one (R) along the
different paths labeled
 1 and 2, corresponding to clockwise and
anti-clockwise rotations respectively. Without loss of generality
we assume that each path covers half a circle ($\pi$ radians). Now
we introduce a magnetic field B parallel to the z-axis to our
model. Within the two level approximation, Eq. (\ref{h0}), the
magnetic field affects the parameters of our model.
If tunneling occures through path  1 (L to R)
the wave function of the particle is influenced
by the factor $e^{i \pi\phi/\phi_0}$, where $\phi$ is the
total flux passing
through a closed tunneling path, $\phi_0$ is the flux quantum defined
by $\phi_0=h/|Q|$, and $h$ is Planck's constant.
On the other hand, tunneling through path number  2 (L to R) adds the factor
$e^{-i \pi\phi/\phi_0}$ to the wave function. Since the tunneling through
both paths occurs with equal probability, the tunneling
matrix element $t$ between L and  R is equal to the sum of the hopping
through path 1 and 2 which is
\be
t(\phi)=\Delta_0(e^{i \pi\phi/\phi_0}+e^{-i\pi\phi/\phi_0})/2=
\Delta_0cos(\pi\phi/\phi_0).
\label{t}
\ee
The factor $1/2$ arises from the equal probability of the paths
in which the total path
probability is conserved to be unity.
Then, replacing the tunneling matrix element $\Delta_0$ in
Eq. (\ref{h0}) by $t(\phi)$, the Hamiltonian of an isolated TLS in the presence
of a magnetic field becomes
\be
H_0(\phi)=\frac{1}{2}  \left(
    \begin{array}{cc}
      \Delta & t(\phi)  \\
      t(\phi) & -\Delta  \\
    \end{array}
  \right).
\label{h0phi}
\ee
The next steps to calculate the physical quantities are similar to
those for the
standard isolated TLS's. \cite{phillips,anderson} To do so, one simply
replace $\Delta_0$ by $t(\phi)$ in all of the equations.

According to Ref. (\onlinecite{kettemann}), the electric permittivity $\epsilon$
depends on temperature as well as magnetic flux through the minimal tunneling
splitting, $t_{min}(\phi)$. The maximum value of
the resonant part of $\epsilon$, $\epsilon_{res}$,
is obtained when the lower cutoff of the
excitation $t_{min}(\phi)$ vanishes. This is important when the temperature
is lowered since the relaxational part of $\epsilon$ becomes negligible.
Thus the maximum value of $\epsilon$ occurs at $t_{min}(\phi)=0$, or
equivalently at $\phi=\phi_0/2$. By introducing the experimental values of
B=$0.1$ T (see Fig. 3 in Ref. \onlinecite{kettemann}) and the typical radius
 $r\approx2\times10^{-10}$m  of a tunneling center one can obtain agreement
with the experimental data, if one further assumes
$\phi_0\approx10^{-5}\times h/|e|$.
In other words, using the definition  $\phi_0=h/Q$, we
arrive at $Q\approx10^{5}|e|$.
The charge $Q$ is suggested to be the effective charge of $N\approx10^{5}$
electrons in TLS's which tunnel coherently.
The coherent tunneling could arise from the interactions between the TLS's
which are important in the low temperature regime.

If we consider  coherent motion with the above
mentioned characteristics, we expect
to observe an oscillation with a frequency proportional to $N$
in the ground state energy
and in some other physical quantity of the interacting model. This is
simply understood by considering $\phi_0=h/(N |e|)$ and substituting it
in the energy eigenvalues of Eq. (\ref{h0phi}), leading to
$\pm\frac{1}{2}\sqrt{\Delta^2+\Delta_0^2\cos^2(N\pi\phi/\tilde{\phi_0})}$,
where
$\tilde{\phi_0}=h/|e|$. We will show in the next sections that a special
form of the interactions can produce such evidence for
coherent motion with the specified property.


\section{Interaction models}
In this section we consider an ensemble of TLS's. Each tunneling
center is placed onto a lattice site. The on-site Hamiltonian is
the same as Eq. (\ref{h0phi}), where $2\Delta$ is the  energy
difference between the well bottoms. The tunneling centers (sites)
interact through $H_{int}$. The Hamiltonian of the lattice is then
of the following form, \be H=\sum_{i=1}^{N} H_0^i(\phi) + H_{int}.
\label{h} \ee The coupling between sites ($H_{int}$) arises from
electric dipole-dipole interactions. The magnetic dipole-dipole
interactions, which are set up by the persistent currents of the
tunneling centers are negligible. Typically the magnetic
interaction is of order $10^{-12}$  smaller than the electric
dipole-dipole interaction for a tunneling system, assuming the
average distance between tunneling centers is $10^{-8}$m.

We first consider the following form for the electric
dipole-dipole interaction,
\be
H_{int}=\sum_{i,j} J_{i,j} \sigma_i^z \sigma_j^z \;\;,
\label{zzhint}
\ee
where $J_{i,j}\sim1/|r_i-r_j|^3$ is the coupling of dipole moments.
We have considered Eq. (\ref{zzhint}) as the interaction term in Eq. (\ref{h}),
and searched for oscillations proportional to $N$ in the ground state
energy or in some other quantity arising from $N$ coupled sites.
We have considered
a one-dimensional array of sites and diagonalized the Hamiltonian exactly
to find its low energy spectrum.
First we considered a model in which all couplings are
fixed to a specified value. We did not observe any level crossing
between the ground and first excited states.
\begin{figure}
\epsfxsize=7.5cm \epsfysize=9.5cm {\epsffile{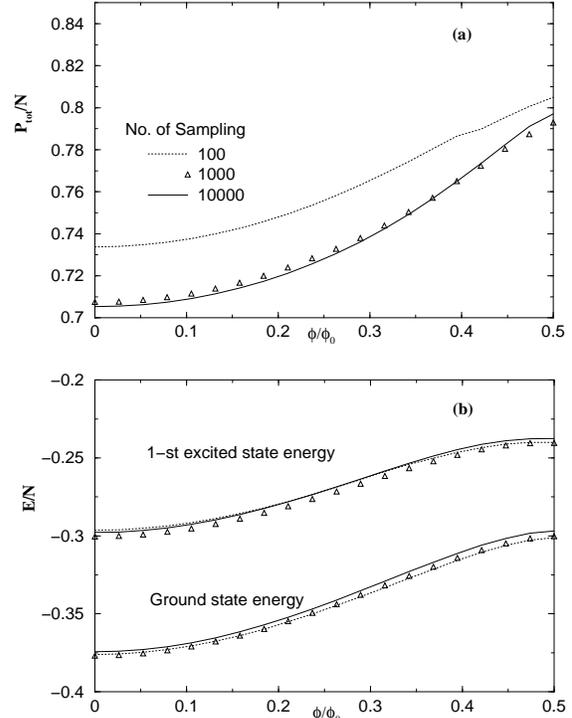}}
\caption{
(a) Dipole moment per site, (b) Ground state and first excited state energy
versus the magnetic flux ratio for N=4 interacting TLS's, Eq. (\ref{zzhint}).
The tunneling centers are
considered to be on a one-dimensional lattice. Data are shown for three different
sampling numbers, 100, 1000 and 10000. It is obvious that the data
converge to the average value in the case of 10000 samplings.
Neither the dipole moment nor the energy levels show any oscillations signaling
coherent motion.
}
\label{fig2}
\end{figure}

In the next step, the coupling constants were chosen random
from a distribution function.
The distribution function $f$ for $\Delta$ and $\Delta_0$ is the same
as in the standard model of TLS's, where $f(\Delta)=f_0$ and $f(\Delta_0)\sim1/\Delta_0$.
Since we consider a one-dimensional array with equal spacing,
$f(r)=1/N$,
the coupling of dipole moments will have the following distribution,
\bea
f(J)=&J_0& J^{-4/3}, \hspace{1cm}
J_1<J<J_2 \nonumber \\
f(\Delta,\Delta_0)&=&\frac{f_0}{\Delta_0}, \hspace{0.8cm}
 \lbrace
    \begin{array}{c}
       0<\Delta<\Delta_{max} \\
       \Delta_{0min}<\Delta_0<\Delta_{0max} ,\\
    \end{array}
\label{p}
\eea
where $f_0$ and $J_0$ are normalization factors.
A system of N=4 sites is studied with random couplings
which are chosen from Eq. (\ref{p}). Typical values for
the parameters are $\Delta_{max}=0.5$K , $0.001$K$<\Delta_0<1$K, and
$0.01$K$<J<1$K. We have computed the energy levels and the dipole moment of the
model for 10000 samples.
Data are plotted in Fig. \ref{fig2}.
We have only computed the data for $0<\phi/\phi_0<0.5$, since
the data are symmetric around $\phi/\phi_0=0.5$, similar to an
even function.\cite{byers}

We consider an external electric field E in the z-direction. Then
the electric field interacts with the dipole moment of the TLS's via
$H_E=-\vec{P}_{tot}\cdot \vec{E}=-E P^z$, which is added to the Hamiltonian.
The dipole moment of the system can be obtained from
$P_{tot}=-\langle0|\partial H/\partial E|0\rangle=
\langle0|\sum_i \sigma_i^z|0\rangle$, where $|0\rangle$ is
the ground state.
The dipole moments vary by a few percent upon changing the
flux ratio (magnetic field), Fig. \ref{fig2}-a, and does not show
any oscillations. In Fig. \ref{fig2}-b, the ground
and first excited state energies do not cross each other
when the flux ratio is changed,
moreover they do not show any oscillations.
We have also considered another domain of the parameters,
long-range interactions resembling a three-dimensional system, and
a distribution of dipole couplings which allows both positive and negative
values. In all cases, similar results were obtained,
which we are not presented.
We conclude that the TLS's with Ising-like interactions of
form of Eq. \ref{zzhint} do
not show any coherent motion.

\begin{figure}
\epsfxsize=7.5cm \epsfysize=9.5cm {\epsffile{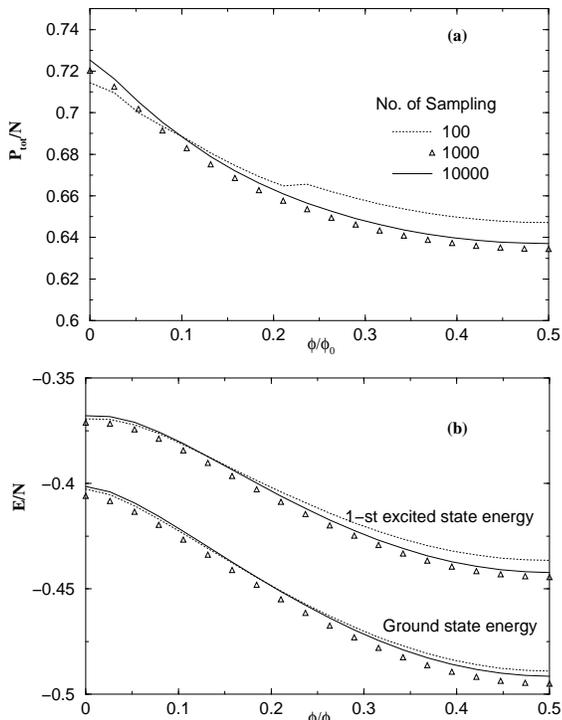}}
\caption{
(a) Dipole moment per site. (b) Ground and first excited state energies
versus magnetic flux ratio for N=4 interacting 3-well potentials.
Tunneling centers are
considered to be on a one-dimensional lattice. Data are shown for three different
sampling numbers, 100, 1000 and 10000. The data
converge to the average value for 10000 samplings.
Neither the dipole moment nor the energy levels show any oscillations
that would indicate
coherent motion.
}
\label{fig3}
\end{figure}

It has been argued that it is not possible to observe coherent
motion in a double-well potential, since a particle tunneling
through either paths  1 or 2 from one well arises at a unique
point which is the other well. \cite{fulde,ahn} The tunneling
entity may tunnel either coherently or randomly to reach the final
configuration. Then the same configuration of some interacting
TLS's is obtained regardless of the relative motion in the TLS's.
One way to overcome this problem is to introduce a 3-well
potential (3WP). \cite{ovchinikov} If we label the 3 wells as  w1,
w2 and   w3, then clockwise or anti-clockwise tunneling from  w1
leads to w2 or  w3, respectively. Then we expect to observe
coherent motion in such a 3WP model.

To study such a model we consider a potential similar to
Fig. \ref{fig1} extended to include  3 wells. The wells are located in a circular
path with equal angle $2\pi/3$. The height of  w1 is $+\Delta$ and of
w3 is $-\Delta$, each  measured from that of  w2. The tunneling matrix element
between the wells is taken to be equal to $\Delta_0$ for simplicity.
By applying a magnetic field to the system the tunneling matrix element is
modified by the factor $e^{i 2\pi\phi/3\phi_0}$.
We define the hopping parameter $t_3=\Delta_0e^{i 2\pi\phi/3\phi_0}$
and $t_3^*$ to be
the tunneling matrix elements between two neighbouring wells in the clockwise
and counter-clockweise directions, respectively.
We then write
the Hamiltonian of a single well as
\be
H_{3,0}(\phi)=\left(
    \begin{array}{ccc}
      \Delta & t_3(\phi) & t_3^*(\phi)  \\
      t_3^*(\phi) & 0 & t_3(\phi)  \\
      t_3(\phi) & t_3^*(\phi) & -\Delta  \\
    \end{array}
  \right).
\label{h30}
\ee
The wells are coupled to each other via an electric dipole-dipole
interaction similar to Eq. (\ref{zzhint}),
\be
H_{3,int}=\sum_{i,j} J_{i,j} m_i^z m_j^z
\label{h3int}
\ee
\noindent
where $m_i^z$ is the z-component of the dipole moment at site $i$. In the
matrix representation, it is a $3\times3$ diagonal matrix whose diagonal elements
are 1, 0 and -1 respectively.

We have performed  similar computations using this interacting
 3WP model.
First, we studied the model using constant parameters, and then
we extended the model to allow the coupling constants to be random,
with distributions according to Eq. (\ref{p}).
In the deterministic case, fixed value of parameters, we did not observe
any level crossing between the ground and first excited states, moreover,
the dipole moment does not show any oscillations
which might indicate collective
phenomena.

In Fig. \ref{fig3} we plotted the dipole moment per site and
the first two low energy levels of N=4 interacting 3WP's. The
data have been computed with the same parameters used in Fig. \ref{fig2}.
The dipole moment versus magnetic flux ratio in Fig. \ref{fig3}-a does not
show any oscillations and we do not observe any level crossing
in the first two energy levels, Fig. \ref{fig3}-b. The same conclusion
is then obtained
that the 3WP model with the interactions presented in
Eq. (\ref{h3int}) does not show evidence for coherent motion.

We thus conclude that
other interaction forms must be present in the model in order to lead to
coherent motion.
Actually, in both
Eqs. (\ref{zzhint}) and (\ref{h3int}), these types of interaction do not depend
upon
the relative motion of the particles in different tunneling centers because
there is no information regarding the tunneling path present
in the Hamiltonian.
These interactions only depend upon the final configuration of
the system and are not able to trace the relative motion of different
entities.
To overcome this difficulty, we will introduce an interaction in the next
section which depends upon $|\theta_i-\theta_j|$, where $\theta_i$ is the
angle defined in Fig. \ref{fig1}. In this case, the interaction contains
information regarding the paths of motion, and more precisely regarding
the relative
motion of different tunneling entities. We will show that a phase with
coherent motion appears for  $N$ coupled TLS's. Surprisingly, this
occurs over a range of parameters, in agreement
with the observed experimental data.

\section{The Modified interaction Hamiltonian}

\subsection{One dimensional model}
In the last section we studied various forms of the interaction
that might enable us to represent the coherent motion as a collective phenomena.
The coherent motion
in the context of a tunneling model is a phase in which all
of the particles in each
potential contribute to the overall tunneling process coherently.
The classical
picture of such a phase is the simultaneous motion of all of the
particles in the same direction, i.e.,
clockwise or anti-clockwise
on a circular path.
To identify such a motion for a particle in the potential of Fig. \ref{fig1},
we define an angle $\theta$ which is measured from the x-axis
by the projection of the dipole moment onto the xy-plane.
When a particle tunnels from one well to the other one, the projected
vector traces a semi-circular path.
In this picture the important quantity is the relative angle
$|\theta_i-\theta_j|$ between pairs of particles in different
tunneling centers. Since the magnitude of this difference is meaningful
modulo $2\pi$, we may define $\cos(|\theta_i-\theta_j|)$ as
being proportional to the
interaction term. Then in the classical picture the interaction
Hamiltonian would be
\be
H^c_{int}=\sum_{i,j} J_{i,j} \cos(|\theta_i-\theta_j|),
\label{hc}
\ee
where $J_{i,j}$ defines the strength of the interaction.
To arrive at the quantum picture, which
is important at low temperatures, we consider the
dipole moment of the particle as an operator acting on the $i$th site.
If we define the projection of the dipole moment onto the xy-plane as
$\vec{p}_{\perp, i}=\sigma_i^x \hat{x}+\sigma_i^y \hat{y}$, then
the $\cos$-term is proportional to the inner product of the
projected vectors,
$\cos(|\theta_i-\theta_j|)\propto\vec{p}_{\perp, i}\cdot \vec{p}_{\perp, j}$.
We use the Pauli matrices, since the on-site Hamiltonian,
Eq. (\ref{h0phi}) has  also been written using this notation.
Thus, the quantum version of Eq. (\ref{hc}) is written in
the following form,
\be
H^q_{int}=\sum_{i,j} J_{i,j} (\sigma_i^x \sigma_j^x +
\sigma_i^y \sigma_j^y).
\label{hq}
\ee
\noindent
Then, the total Hamiltonian of interacting system is
\be
H_t=\sum_{i=1}^N H_0^i(\phi) +
\sum_{i,j} J_{i,j} (\sigma_i^x \sigma_j^x + \sigma_i^y \sigma_j^y).
\label{ht}
\ee
\noindent
Before starting to study the properties of $H_t$, we would like to
mention the most important difference between Eqs. (\ref{zzhint})
and (\ref{hq}). The Hamiltonian of Eq. (\ref{zzhint}) has the discrete
$Z2$ symmetry  whereas Eq. (\ref{hq}) has the continous $U(1)$ symmetry.
Moreover,  we are now able to trace the
tunneling path using the continuous symmetry, and  to define the phase
of coherent motion.

We will first study the properties of $H_t$ using constant
parameters, and then we shall consider the effects of random couplings
subject to a parameter distribution function.
\begin{figure}
\epsfxsize=7.5cm \epsfysize=9.5cm {\epsffile{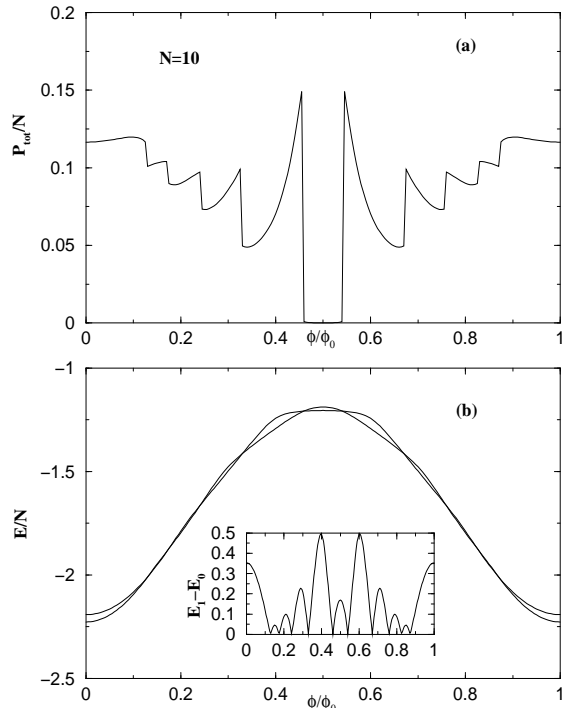}}
\caption{
(a) Dipole moment per site. (b) Ground and first excited state energies
versus the magnetic flux ratio with N=10 interacting TLS's
from Eq. (\ref{ht}).
The inset in (b) is the difference ($E_1-E_0$) of the first excited
and ground state energies, which illustrate more clearly the
level crossings.
There are 10 points of level crossings and exactly at these points a sharp
change in the dipole moment appears.
We have treated a one-dimensional
array of tunneling centers with the  parameters
$\Delta=0.2$, $\Delta_0=3$, and $J_{i,j}=J=1$.
}
\label{fig4}
\end{figure}
Let us consider a one-dimensional array of tunneling centers with
the total Hamiltonian defined by $H_t$. We shall examine two quantities,
the dipole moment, induced dipole moment, as we have defined it earlier, and
the two lowest-lying energy levels.
In Fig. \ref{fig4} we have plotted the dipole moment per site and the
first two energy levels of $N=10$ interacting TLS's. The following
couplings have been used to observe the present behavior,
$\Delta=0.2$, $\Delta_0=3$, and $J_{i,j}=J=1$. The dipole moment is induced
by the external electric field when this effect is allowed
due to the asymmetry in the height of the wells, $\Delta\neq0$.
Thus the amplitude
of the dipole moment is proportional to $\Delta$, which is small.
Moreover we observed $N$ oscillations in the dipole moment,
which is the signature
of  coherent motion. Exactly at the position of a sharp change in the
dipole moment, a level crossing between the first two low-lying
states occurs, as has been plotted in Fig. \ref{fig4}-b.
\begin{figure}
\epsfxsize=7.5cm \epsfysize=9.5cm {\epsffile{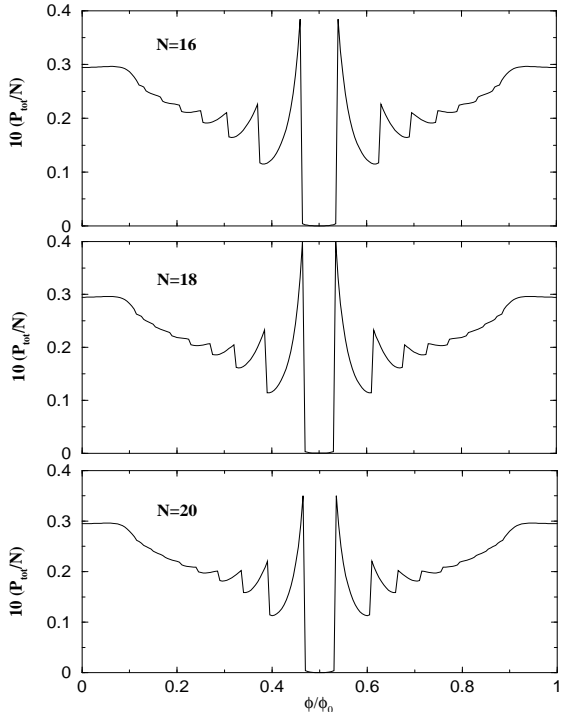}}
\caption{
(a) Dipole moment per site magnified by 10 versus the magnetic flux ratio with
N=16, 18, and 20 interacting TLS's.
We have considered a one-dimensional
array of tunneling centers with the parameters
$\Delta=0.05$, $\Delta_0=3$, and $J_{i,j}=J=1$.
}
\label{fig5}
\end{figure}

To show the level crossings more clearly, we have shown the
difference between the first excited state energy $E_1$ and the
ground state energy $E_0$  in the inset of Fig. \ref{fig4}-b.
These observations confirm that the model introduced in
Ref. \onlinecite{kettemann} can be the effective model for the
interacting system defined in Eq. (\ref{ht}).
In this sense, the oscillations in the dipole moment or the number of
level crossings between the first two low-lying states, represent the
number of tunneling entities which move coherently. Therefore,
the Hamiltonian of Eq. (\ref{ht}) describes a system of $N$ interacting
TLS's in the  coherent motion phase. In the following, we will discuss
the range of parameters over which this phase is observed.
The specified couplings are in accordance with the low temperature regime,
where the collective phenomena of cold glasses has been observed.
By changing the parameters, the phase may appear at higher or
lower temperatures.

To see the dependence of the dipole moment on the number of tunneling centers,
we
have plotted this quantity for $N=16, 18, 20$ in Fig. \ref{fig5}.
In this
figure, we have fixed the couplings to
$\Delta=0.05$, $\Delta_0=3$, and $J_{i,j}=J=1$.
In Fig.(\ref{fig5}), we magnified the plot tenfold to more clearly observe
the oscillations.
We still observe N oscillations in
$P_{tot}/N$ versus magnetic flux ratio but the height of the discontinuities
has been reduced. Moreover the spacing between these  points is smaller for
larger N. This is due to the size dependence of the extensive physical
quantities. When N is increased, the difference between the energy levels
and the distance between two level crossing points are both reduced. Then, the
level crossing occurs for two states which are nearly degenerate.
In this case, the states are similar to each other, and their
dipole moments differ only slightly, so we do
not observe a pronounced discontinuity. In the  thermodynamic
limit, these two states  become degenerate. Thus we would like
to stress that the phenomena we have predicted are mesoscopic
effects which occurs for finite N and not in the
thermodynamic limit $N \rightarrow \infty$. This is surprisingly in
agreement with the effective theory of TLS's, \cite{kettemann}
in which the number
of interacting TLS's is assumed to be $\sim 10^5$.

\begin{figure}
\epsfxsize=7.5cm \epsfysize=9.5cm {\epsffile{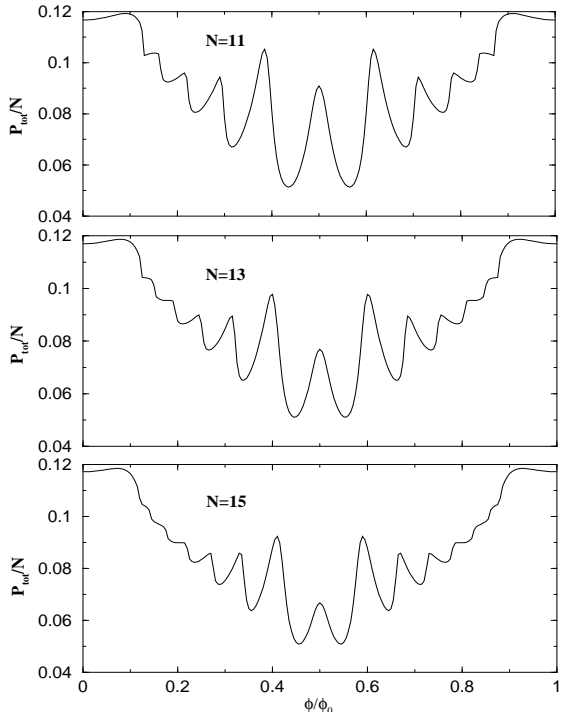}}
\caption{
Dipole moment per site versus the magnetic flux ratio for
odd size (N=11, 13, 15) interacting TLS's.
The coupling constants of the one-dimensional Hamiltonian are
$\Delta=0.2$, $\Delta_0=3$, and $J_{i,j}=J=1$.
}
\label{fig6}
\end{figure}

The other feature which supports that the notion that the observed
phenomena are mesoscopic, is the even and odd dependence of the
dipole moments.\cite{ahn} We have so far presented the quantities
for only even values of N. The results for odd values of N are
plotted in Fig. \ref{fig6}. The common feature is the number of
oscillations in the dipole moment, which is exactly equal to N. We
have shown this fact for $N=11, 13, 15$, with the same couplings
used in  Fig. \ref{fig4}. The main difference between even and odd
N appears in the form of the oscillations in the dipole moment.
For the even N case, at the locations of each level crossing, the
ground state changes from one state to another, leading to a
discontinuity in the dipole moment. The case of odd N values is
different. Hence the oscillations arise from a smooth change in
$P_{tot}/N$. In particular, the oscillations arise from the beat
frequencies of the nearly degenerate ground and first excited
states, without any level crossing. By changing the magnetic flux
ratio the ground and first excited states switch positions,
  due to the alternation in their relative energy differences.
Each time the ground state
comes close to first excited state, a peak in the dipole moment occurs.

The origin of this different behavior for even and odd N is related to
the difference in the degeneracies
of the ground state at the magnetic flux ratio, the level crossing points
for even N, and the level beatings for odd N, respectively.
The degeneracy is two fold, and
can be labeled by the eigenvalues of a parity.
Suppose in $H_t$ we set $\Delta=0$,
the symmetric TLS or in the absence of external electric field. Then the
Hamiltonian $H_t(\Delta=0)$ is invariant under the parity operation
$\Pi(\hat{z})=-\hat{z}$,
which defines the $Z2$ symmetry. Then each eigenstate of the Hamiltonian will
be an eigenstate of $\Pi$.
Since $\Pi^2=\mathcal{I}$, the identity operation,
the eigenvalues of $\Pi$ are $+1$ or $-1$.
When the electric field is turned on, $\Delta\neq0$ the $Z2$ symmetry is
broken and we expect it to remove the degeneracy at the level
crossing points. Actually, this is true for odd N, for which the degeneracy is
removed by an electric field as in the linear stark effect, there is no
level crossing, and the dipole moment changes continuously.
This means the first two low-lying eigenstates of the Hamiltonian with
odd N have different $Z2$ parity eigenvalues, whereas they have the same
parity for even N.
The degeneracy
still remains at a level crossing for even N, and leads to
discontinuous changes in the dipole moment. Thus we should search for another
symmetry which might be responsible for the degeneracy present for even N.
We define the mirror image of a one-dimensional array to its center by
$\Omega(i)=N-i+1$ where $i$ is the site label, and $N$
is the length of array.
Again $\Omega^2= \mathcal{I}$, which defines the eigenvalues
$\omega_{1(2)}=+1(-1)$
respectively. Since $\Omega$
commutes with $H_t$, all eigenstates of the Hamiltonian are also
eigenstates of $\Omega$. Thus, we conclude that the first two low-lying
eigenstates of the even N  Hamiltonian  have different $\Omega$ parity.

\begin{figure}
\epsfxsize=7.5cm \epsfysize=9.5cm {\epsffile{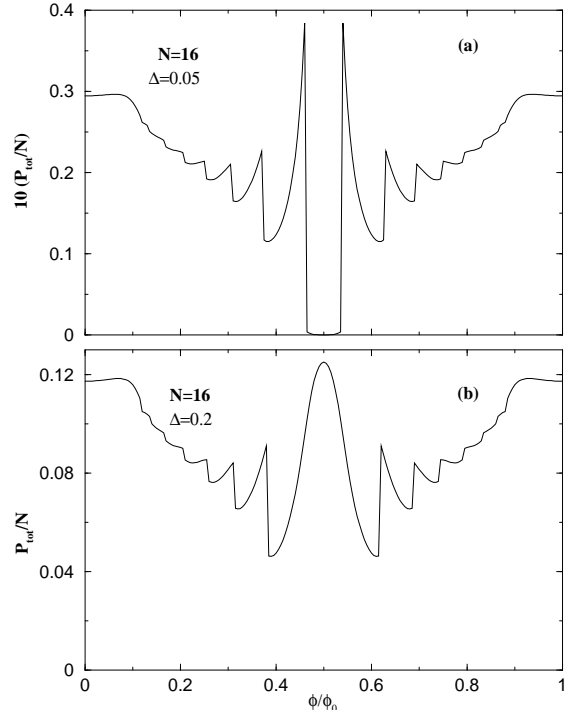}}
\caption{
The dependence of dipole moment per site on the asymmetry
parameter versus magnetic flux ratio of N=16 interacting TLS's.
(a) All TLS's contribute to the coherent motion, where $\Delta=0.05$.
(b) The effective number of TLS's in  coherent motion is reduced
to 15 when $\Delta=0.2$.
The other coupling constants are
$\Delta_0=3$ and $J_{i,j}=J=1$.
}
\label{fig7}
\end{figure}

Now we discuss on the range of parameters which gives
the phase with coherent motion. The main coupling constants in
our model are $\Delta_0$ and $J$. Since the value of $\Delta$ corresponds to the
asymmetry of the wells, it is considered to be much smaller than the other
parameters. At large  $\Delta$ value,
the  energy difference of the two wells is too high to allow tunneling, so
some of the particles will be frozen
in the lowest well of their own tunneling center, and do not contribute
to the coherent motion. Then the probability of having
all particles moving coherently will be reduced. This can be shown
by computing the dipole moment for different values of $\Delta$.
In Fig. \ref{fig7}, we plotted the dipole moment of a one-dimensional
 array of $N=16$ TLS's with different asymmetries, $\Delta=0.05$ in
part (a) and $\Delta=0.2$ in part (b). We observe 16 peaks in
Fig. \ref{fig7}-a, but this number is reduced to 15 in Fig. \ref{fig7}-b,
which shows that only 15 of the tunneling centers are effectively moving
coherently. With increasing $\Delta$, this number decreases.

So by considering a small value for $\Delta \sim 1/N$ for a mesoscopic
size (N) system, the ratio of the other two parameters ($\lambda=\Delta_0/J$)
controls the behavior of the system.
If we fix $J=1$, then $\Delta_0$ is the tunable parameter.
We have found two different phases in our model.
When $\lambda < \lambda_1$ the number of level crossings is less than N.
In this phase, just some of the tunneling centers
contribute to the coherent motion, and for very small values of $\lambda$
the tunneling probability is negligible, and consequently, there is no
coherent motion. This occurs when the interaction coupling ($J$)
is very strong and causes the system to be frozen into a ground state which
is only defined by the interaction term.
In the other phase, when $\lambda > \lambda_1$, the model shows
N level crossings between the first two low-lying levels and consequently
the dipole moment
oscillates with a frequency proportional to N. This phase corresponds
to coherent motion. By increasing $\lambda$, the model behaves
similar to that for independent TLS's, since $\Delta_0$ is the
strongest parameter
in the Hamiltonian, and hence the interaction term becomes negligible.
Practically the coherent motion phase becomes observable in the range
of parameters $\lambda_1 < \lambda < \lambda_2$.
We have found that for one-dimensional model, $\lambda_1(1D)=3$ and
for a two-dimensional lattice, it is $\lambda_1(2D)\simeq6$.
So we assume $\lambda_1(3D)\sim 10$ for a three-dimensional model.
If we choose a typical value of $\Delta_0=1$K, then the
strength of the dipole-dipole interaction ($J$) would have to be $J=100$mK
in order to
observe a collective phenomenon.
This estimated value of $J$ obtained from our model is surprisingly in agreement
with the expected value of the dipole-dipole interaction, assuming
the average distance between two tunneling centers is $R=10^{-8}$m and
the average dipole moment of each TLS is $P=2|e|\times10^{-10}$m.

\subsection{Two dimensional model}

In this subsection we present the results of our
 calculations for the dipole moment
of the Hamiltonian defined in Eq. (\ref{ht}) on a two dimensional
square lattice. We only considered the nearest neighbour (n.n)
interaction. The coupling constants are fixed
to the values  $\Delta=0.2$, $\Delta_0=6$, and
$J_{i,j}=J=1$. The dipole moment per site on the finite 2D model
is plotted in Fig.\ref{fig8} for lattice sizes of
$N=3\times3\;,\; 4\times3\;,\; 4\times4$ and $4\times5$.
The expected oscillations
in the dipole moment are observed in all cases. The oscillations
are more apparent for the smaller $N$ (=9, 12) cases. As we discussed
previously, by increasing $N$ the difference between the energy levels
is decreased, and
consequently the magnitude of the oscillations in the dipole moment is reduced.
Moreover, we still distinguish the different types of oscillations for
odd and even system sizes, which is clear by comparing Figs. \ref{fig8}-a
and \ref{fig8}-b for $N=9$ and $12$, respectively.
In the case of odd $N$, the oscillations do not lead to a
discontinuity in the dipole moment, whereas  it is discontinuous
for even $N$.
Our data for the two-dimensional system confirms that the proposed model
is able to obtain a coherent motion in higher dimensions. This will
be justified by considering long-range interactions in the next subsection.

\subsection{Long range interactions}

We now consider the effect of including long-range (l.r.) interactions
in our model. Since we are attempting to describe a glassy system,
the positions of
the TLS's are not limited to the sites on a regular lattice.
Hence, each tunneling
center  may have many neighbors. In this respect, we have studied
long-range interactions in both one- and two-dimensional models.
We have calculated the dipole moment and the low-energy spectrum in order to
compare them with the results for nearest-neighbor (n.n.) interactions
studied in the last two subsections.
\begin{figure}
\epsfxsize=7.5cm \epsfysize=9.5cm {\epsffile{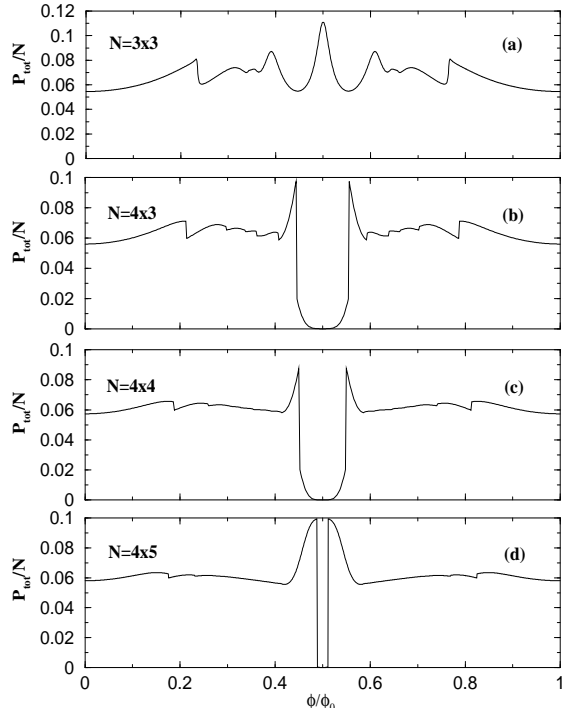}}
\caption{
Dipole moment per site for the two-dimensional model
versus the magnetic flux ratio.
The lattice sizes are (a) N=9, (b) N=12, (c) N=16, and (d) N=20.
The coupling constants are
$\Delta=0.2$, $\Delta_0=6$, and $J_{i,j}=J=1$, where
the interaction is only between nearest neighbors.
}
\label{fig8}
\end{figure}

\begin{figure}
\epsfxsize=7.5cm \epsfysize=9.5cm {\epsffile{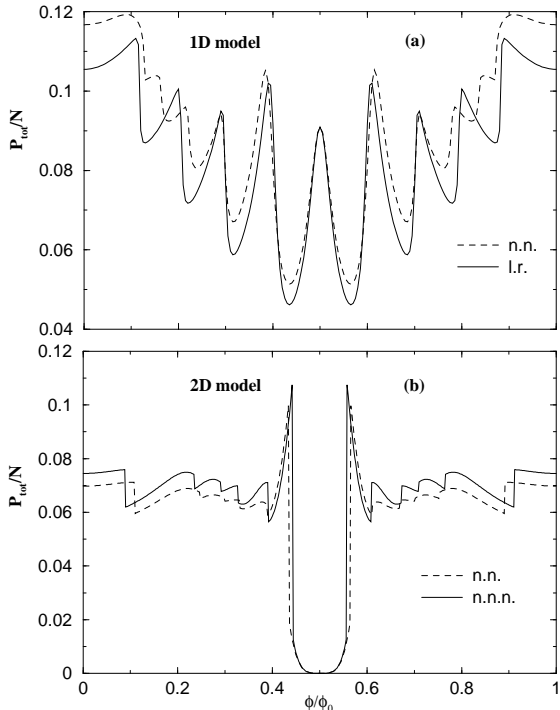}}
\caption{
The effect of long-range (l.r.) interactions on the
dipole moment per site for one- and two-dimensional lattices
versus the magnetic flux ratio.
(a) The 1D lattice size is N=11 and the coupling constants are
$\Delta=0.2$, $\Delta_0=6$, and $J_{i,j}=J=1$ for n.n.,
and $J_{i,j}=1/(r_{i,j}^3)$ for l.r. interactions.
(b) The 2D model is a $4\times3=12$ square lattice where
$\Delta=0.2$, $\Delta_0=5$, and $J_1=1$ for n.n., and
$J_2=0.125$ for next n.n. interactions, the latter representing
a long-range interaction.
}
\label{fig9}
\end{figure}

In a one-dimensional array of long-range interacting TLS's,
the indices $i$ and $j$ run over
all lattice points. Since the interactions originate from dipole-dipole
couplings, the strength of the interaction between sites
$i$ and $j$ should be equal to
$J_{i,j}=J/(r_{i,j}^3)$. We have plotted the resulting dipole moment per site
for an array of size $N=11$ in Fig. \ref{fig9}-a. In this plot, we have
presented  data both for n.n. and l.r. interactions. The nearest-neighbor
coupling strength is taken to be $J=1$ in both cases.
There is no qualitative change in
the behavior of the dipole moment versus the magnetic flux ratio.
This is due to the ordering in the coherent motion phase.
Suppose that (i, j) and (j, k) are two n.n. sites. If the system is in the
coherent motion phase, the relative angles between n.n. sites is constant,
$|\theta_i-\theta_j|=c_{ij}\;,\; |\theta_j-\theta_k|=c_{jk}$.
By turning on the l.r. interaction between (i, k), the constraint
$|\theta_i-\theta_k|=c_{ik}$ is imposed on the system,
consistent with the previous system configuration, as seen by noting
that $c_{ik}\equiv c_{ij}+c_{jk}$. Thus, adding l.r. interactions
does not destroy the coherent motion.
Moreover, the oscillations are more pronounced
in the case of l.r. interactions,
and also the distance between two peaks becomes approximately equally spaced.
Then we observe more clearly the oscillations
in the dipole moment with a frequency proportional to the lattice size.
When we consider l.r. interactions to all neighbor in the lattice, the system
resembles a three dimensional one, since each site has
many neighbors. This adds for the support to the notion that this
model may describe  three-dimensional coherent motion.
In Fig.\ref{fig8}-b we present our results for
 a similar situation
for a two-dimensional lattice. The lattice size is $N=4\times3=12$,
where a TLS exists at each lattice point.
For simplicity we have only considered the
next nearest-neighbor (n.n.n.) interactions, using it to represent
a long-range interaction, since it
does not change the generality.
In this case, the n.n. coupling is $J_1=1$ and the n.n.n. coupling
is $J_2=0.125$, which
are supposed to behave as $1/r^3$.
As discussed for the 1D case, there are no qualitative changes in the
results. In comparison with the n.n. results,
the oscillations in the dipole moment are observed to be clearer and more
equally spaced.

\subsection{Random couplings}

So far, we have studied the Hamiltonian defined in Eq. (\ref{ht}) for
one- and two-dimensional lattices with n.n. and l.r. interactions.
In all of these cases we have considered a deterministic model,
in which all of the couplings were constant and homogeneous for
all lattice points.
Since we are attempting to describe a phenomena in a glassy media, we
expect to have a distribution of coupling constants.

\begin{figure}
\epsfxsize=7.5cm \epsfysize=9.5cm {\epsffile{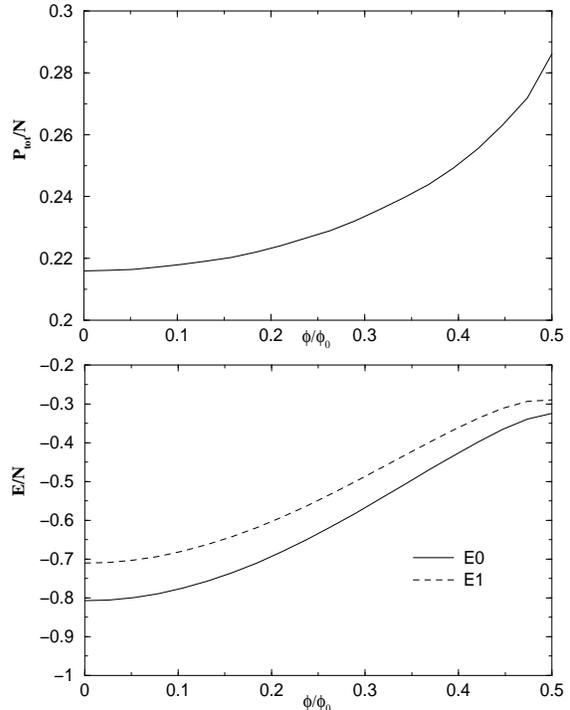}}
\caption{
(a) Dipole moment per site and (b) two lowest energy
levels versus the magnetic flux ratio.
The data were obtained using the random couplings defined
in Eq. (\ref{p}) with the ranges of parameters
$0<\Delta<0.2$, $0.05<\Delta_0<3$, and
$0.05<J<1$. The averaging was done over 10000 samples
for a system of size $N=4$.
Both parts of this figure show no oscillations, and hence do not
provide support the notion of collective behavior.
}
\label{fig10}
\end{figure}

The first model is to take the same distribution
that has been used to treat
non-interacting TLS's. The distribution function
is written in Eq. (\ref{p}) as $f(\Delta,\Delta_0)$, and
for a one-dimensional regular lattice, the distribution of
the dipole-dipole interactions is $f(J)$ in the same equation.
The distribution for a three dimensional lattice
is calculated to be $f(J_{i,j})\simeq J_{i,j}^{-2}$ by assuming
$J_{i,j}\sim1/r_{i,j}^3$. Now let us limit our study
to the case defined in Eq. (\ref{p}) for a one-dimensional
array of TLS's. We have calculated the dipole moment and the
first two low-lying energy levels of a system with $N=4$ interacting TLS's
where the ranges of parameters are  $0<\Delta<0.2$, $0.05<\Delta_0<3$, and
$0.05<J<1$. We have calculated the average values for $10000$ samples
and plotted them in Fig.\ref{fig10}.
Since the data are symmetric around $\phi/\phi_0=0.5$,
we presented them only for $[0,0.5]$.
As shown in Fig.\ref{fig10}-a,
the dipole moment does not demonstrate any oscillations, but instead
shows monotonic behavior over the domain model.
Moreover the first two energy levels  exhibit neither
a level crossing nor a level frequency beating.
This calculation verifies that the model with
this type of distribution function does not describe coherent
motion. Although we presented the data only for a special range of parameters,
we did not observe any qualitative changes in the results by altering the
range of parameters.
We note that the type of  distribution function
$f(\Delta,\Delta_0)$ defined in Eq. (\ref{p})
follows from the assumption that the TLS's in  glasses
are considered to be isolated.
Thus we argue that for interacting TLS's the
distribution  function should be modified!

\begin{figure}
\epsfxsize=7.5cm \epsfysize=9.5cm {\epsffile{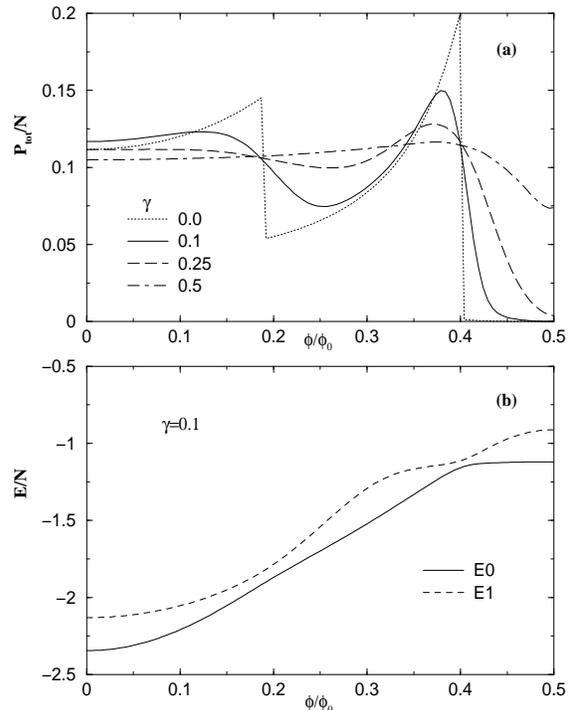}}
\caption{
(a) Dipole moment per site versus the magnetic flux ratio with
Gaussian randomness for an $N=4$ one-dimensional array
of interacting TLS's.
The width $\tilde{a}$ of the Gaussian distribution
satisfies
$\tilde{a}=\gamma \bar{x}$, where $\bar{x}$ is the average value of the
random variable. Data are presented for $\gamma=0, 0.1, 0.25, 0.5$.
For small values of $\gamma$ the oscillations in the dipole moment are
strong and clearly evident, providing evidence
for coherent motion.
With increasing $\gamma$, the model loses its evidence for collective behavior.
(b) Ground and first excited state energies of the model
for $\gamma=0.1$ which show the level frequency beating, a sign of
coherent motion.
}
\label{fig11}
\end{figure}

Hence, we choose a Gaussian distribution for the parameters in our
model, \be g(x)=\frac{1}{\tilde{a} \sqrt{2\pi}}
e^{-\frac{(x-\bar{x})^2}{2 \tilde{a}^2}} \label{gauss} \ee where
$\bar{x}\equiv\langle x \rangle$ is the average value of the
variable $x$ and $\tilde{a}\equiv\sqrt{\langle x^2 \rangle -
\langle x \rangle ^2}$ its width. Now we consider a Gaussian
distribution for all of the parameters in our model. In
Fig.\ref{fig11}-a, the dipole moment of a one-dimensional array of
TLS's with $N=4$ is plotted versus the magnetic flux ratio for
four different values of the  parameter $\gamma$. This parameter
defines the variance of the distribution by $\tilde{a}=\gamma
\bar{x}$. The average values of parameters are $\bar{\Delta}=0.2$,
$\bar{\Delta_0}=3$, and $\bar{J}=1$. We have computed the average
values of the physical quantities over $10000$ samples. In this
notation, $\gamma=0$ is the deterministic case in which there is
no random variable, leading to very  sharp oscillations in the
dipole moment. As discussed in the last sections, the dipole
moment behaves discontinuously at the level crossing points when
$N$ is even. The total number of oscillations is four, with two in
the region $\phi/\phi_0 \in [0,0.5]$ (Fig.\ref{fig11}-a), and two
in the $\phi/\phi_0 \in [0.5,1]$ region, which are not presented
to save computation time. For $\gamma=0.1$, the variance is equal
to 10 percent of the average value and still the oscillations in
the dipole moment are clear, confirming coherent motion. For
$\gamma=0.1$ we have also presented the two lowest energy levels
in Fig.\ref{fig11}-b, in which the effect of level frequency
beating instead of level crossing can be seen. Since we included
random variables in the model, we do not expect  sharp level
crossing. But the oscillations with frequency proportional to $N$
in the energy levels show the dependence of the energy on the
number of interacting TLS's, suggesting collective behavior. If
variance is less than or equal to 10 percent of the average value,
we still will observe evidence for coherent motion. By increasing
$\gamma$, the amplitude of the oscillations is reduced, and for
$\gamma=0.5$, there are almost no oscillations. In other words, a
wide distribution of randomness destroys the collective behavior.
Hence, we have shown that the proposed model exhibits coherent
motion provided that the distribution of the coupling constants is
sufficiently narrow.

\section{Conclusion}
We have attempted to find an appropriate interaction between TLS's
in order to explain the experimental evidence for collective
phenomenon in multi-component glasses at ultra low temperatures.
Our first attempt was to distinguish the features of double well
and triple well potentials in the presence of Ising-like
interactions. The Ising interactions are intended to simulate the
z-component of the dipole-dipole interactions between two wells
when one uses a distribution of  exchange couplings to simulate
different dipole moment orientations. Our calculations confirm
that none of the two- and three-well models with Ising-type
interactions can provide evidence for a coherent motion phase.
This is true for both fixed and random couplings. The crucial
point arises from the symmetry of the interaction. The Ising
interaction has a discrete symmetry, whereas a continuous symmetry
is required in order to trace the path of the tunneling process
for the simulation of coherent motion.

Second, we considered an XY interaction  proportional to
the $\cos(|\theta_i-\theta_j|)$, where $\theta_i$ is the angle of
the dipole moment
projected onto the xy-plane (see Fig.\ref{fig1}). Thus the interaction is
sensitive to the relative angle of the dipole moments in
two coupled TLS's. In the
coherent motion phase, the relative angle of each coupled TLS's is constant, so
the ensemble of TLS's demonstrates a collective phenomenon in which all of the
particles tunnel in the same direction (clock- or anti-clockwise).
One may treat at this phenomenon effectively by a single TLS with a charge
equal to the sum of the charges of all of the correlated TLS's.
The XY interaction has a continuous $U(1)$ symmetry which properly traces out
the path of tunneling and simulates the coherent motion.
We have studied this model on one- and two-dimensional lattices in which the
tunneling centers are restricted to the lattice points. We have predicted the
coherent motion phase in a range of coupling constants
by calculating the dipole moments and the low-energy levels of the coupled
TLS's.

The effects of long-range interactions were studied in a model
exhibiting the expected clearly and
equally spaced
oscillations in the dipole moment.
By definition, in the coherent motion phase the relative angle between
coupled TLS's is constant.
Adding l.r. interactions imposes constraints between non-nearest
neighbors which are effectively satisfied by imposing a n.n. constraint.
Thus l.r.
interactions do not destroy the coherent motion phase.
Moreover, results obtained with l.r. interactions resemble those obtained in
higher spatial dimensions,
which suggests that our model  may be relevant for the 3D case as well.
The glassy properties of our model were examined by introducing random couplings.
We have found that the form of the random distribution  depends
upon the interacting or non-interacting model. If we assume the
distribution of isolated TLS (Eq. (\ref{p})), for the
interacting model, we will not observe any evidence for the
coherent motion phase.
Thus we have considered a Gaussian distribution for all of the parameters
in our model. When the width of the Gaussian distribution
is narrow, we still observe the
oscillations in the dipole moment representing collective behaviour.
For a wide distribution of parameters  the
coherent motion phase is destroyed.

Finally we conclude that the XY-like interactions between TLS's,
combined with the Aharanov-Bohm effect of the single particle
Hamiltonian, allows us to construct a successful model exhibiting
the important features suggestive of the  collective phenomenon of
coherent motion in low temperature glasses. There are still many
aspects of the model such as the density of states as well as
finite temperatures which should be investigated in future
studies.

\section{Acknowledgement}

I would like to express my deep gratitude to P. Fulde who introduced
to me this topic and for
valuable comments, discussions and a careful study of the manuscript.
I would also like to thank K.-H. Ahn, A. Bernert, S. Kettemann,
I. Peschel and
S. Pleutin for fruitful discussions. Many thanks to R. Klemm for his
comments on the manuscript. I would like also to thank A. A. Ovchinnikov
for making his calculation available before publication.


\end{multicols}

\begin{references}

\bibitem{esquinazi}
Tunneling Systems in Amorphous and Crystalline Solids, edited by
P. Esquinazi (Springer, Heidelberg, 1998).

\bibitem{phillips}
W. A. Phillips, J. Low Temp. Phys. {\bf 7}, 351 (1972).

\bibitem{anderson}
P. W. Anderson, B. I. Halperin and C. M. Varma, Philos. Mag. {\bf 25},
1 (1972).

\bibitem{enss}
C. Enss and S. Hunklinger, \prl {\bf 79}, 2831 (1997).

\bibitem{yu}
C. C. Yu and A. J. Leggett, Comments Condens. Matter Phys. {\bf 14},
231  (1988); H. M. Carruzzo, E. R. Grannan and C. C. Yu, \prb
{\bf B50}, 6685 (1994).

\bibitem{kassner}
K. Kassner and R. Silbey, J. Phys. Condens. Matter {\bf 1}, 4599 (1989).


\bibitem{wurger}
A. W\"urger, Z. Phys. {\bf B98}, 561 (1995);
C. Enss, et.al \prb {\bf B53}, 12094 (1996).

\bibitem{kuhn}
R. K\"uhn and A. W\"urger, \prb {\bf B62}, 12069 (2000).

\bibitem{strehlow}
P. Strehlow, C. Enss and Hunklinger, \prl {\bf 80}, 5361 (1998);
S. Hunklinger, et.al Physica {\bf 263-264}, 248 (1999);
S. Hunklinger, et.al Physica {\bf 280}, 271 (2000);
P. Strehlow, et.al \prl {\bf 84}, 1938 (2000).

\bibitem{reijntjes}
P. J. Reijntjes, W. van Rijswijk, G. A. Vermeulen and G. Frossati,
Rev. Sci. Instrum. {\bf 57}, 141 (1986).

\bibitem{kettemann}
S. Kettemann, P. Fulde and P. Strehlow, \prl {\bf 83}, 4325 (1999).

\bibitem{byers}
N. Byers and C. N. Yang \prl {\bf 7}, 46 (1961).

\bibitem{fulde}
P. Fulde, privite communication.

\bibitem{ahn}
K. -H. Ahn and P. Fulde, \prb {\bf 62}, R4813 (2000).


\bibitem{ovchinikov}
A. A. Ovchinnikov, "Three-well centers in glasses", International workshop
on collective phenomenon in the low-temperature physics of glasses,
October 2000, Dreseden-Germany.





\end{references}
\end{document}